\documentclass[pdftex]{nolta}

\usepackage{amsmath}
\usepackage{mathtools}

\Vol{0}%
\No{0}%
\Year{2025}%
\Month{9}%

\title{Statistical properties of homoclinic bursting: an approach using infinite-modal maps toward predicting extreme events}

\AUTHOR*{%
\author{Masaki Nakagawa}{1}\orcid{0000-0001-7584-5970}
}

\AFFILIATE{%
\affiliate{Department of Computer Science and Engineering, \\
Fukuoka Institute of Technology, \\
3-30-1 Wajiro-higashi, Higashi-ku, Fukuoka 811-0295, Japan}{1}
}

\received{}{}{}
\revised{}{}{}
\published{}{}{}

\begin{document}

\begin{abstract} 
Predicting extreme events in nonlinear dynamical systems is challenging due to a limited understanding of their statistical properties.
This study numerically and theoretically investigates the statistical properties of infinite-modal maps arising from homoclinic bursting to predict extreme events. 
The numerical investigation presents bifurcation diagrams, Lyapunov exponents, height probability distributions, and interevent interval probability distributions for infinite-modal maps.
The theoretical analysis derives analytical formulae for these statistical properties using a randomization theory of infinite-modal maps.
Furthermore, a parameter estimation method for infinite-modal maps is proposed, utilizing the derived analytical formula, which enables practical application of the theoretical results.
Finally, the study demonstrates the applicability of the approach in analyzing non-stationary data with time-dependent parameters.
These findings provide a foundation for the prediction of extreme events based on their mechanism.
\end{abstract}


\maketitle 
\thispagestyle{empty}

\section{Introduction}\label{sec:intro}
Extreme events occur in various physical and engineering systems, including turbulence~\cite{Ishihara+2007,Yeung+2015}, weather systems (e.g.\ ENSO)~\cite{Timmermann+2003,Ray+2020}, optical systems (rogue waves)~\cite{Solli+2007,Bonatto+2011}, and nonlinear mechanical oscillators~\cite{Durairaj+2023,Zhao+2023}. 
These events are rare but can have devastating impacts, motivating a wide range of studies.
Recent progress of such studies has been substantial in data-driven prediction~\cite{Farazmand&Sapsis2019,Sapsis2021}.
However, developing analytically tractable and computationally efficient prediction methods remains a significant challenge, especially under conditions of data scarcity.
Bridging this gap is essential for building predictive frameworks based on the mechanisms that generate extreme events.
This study takes a step in this direction by introducing and analyzing a reduced dynamical model arising from homoclinic bursting.

\begin{figure}[tb]
    \centering
    \includegraphics[width=.4\linewidth]{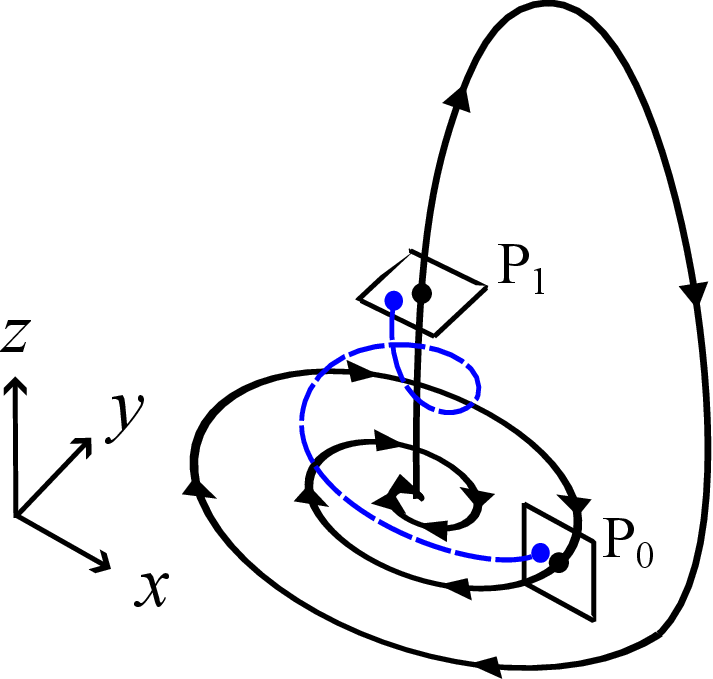}
    \caption{Schematic of a homoclinic orbit to a saddle-focus point.
    A dashed orbit shows an orbit from a cross section $P_0$ to another cross section $P_1$.
    }\label{fig:homoclinic-loop}
\end{figure}

A mechanism-based perspective has shown that specific dynamical structures often underlie the occurrence of extreme events. 
Such structures include slow manifolds~\cite{Fenichel1979,Jones1995}, noise-induced transitions~\cite{Horsthemke&Lefever1984,Forgoston&Moore2017}, and homoclinic orbits~\cite{Shilnikov1965,Gaspard&Nicolis1983}.
Among these, this study focuses on the third mechanism.
Homoclinic orbits to a saddle-focus point {provide} a typical mechanism for bursting, commonly referred to as homoclinic bursting~\cite{Farazmand&Sapsis2019}, as shown in Fig.~\ref{fig:homoclinic-loop}.
This insight has enabled the real-time prediction and suppression of extreme events in numerical settings. 
For example, closed-loop adaptive control that intervenes only when a burst is imminent has been proposed in turbulent shear flows~\cite{Farazmand&Sapsis2019PRE}.
Additionally, related work explored Markov model-based schemes for predicting and mitigating extreme events~\cite{Kaveh&Salarieh2020}.
These approaches typically rely on extensive simulations or numerical optimization, which obscures explicit links between model parameters and observable statistics. 
Consequently, the parameters are challenging to infer quickly from data or reuse in similar other systems.
Moreover, because such methods often depend on large amounts of data, their applicability under data scarcity remains uncertain.

This study employs a two-dimensional dynamical system, particularly the {\it Pacifico-Rovella-Viana (PRV) map}, as a simple surrogate model for extreme events. 
Such a dynamical system universally arises as a Poincar\'{e} return map near a homoclinic orbit and belongs to a class of infinite-modal maps~\cite{Pacifico+1998}.
An infinite-modal map is a map that possesses a countably infinite number of critical points.
Although these maps are often studied in fields such as bifurcation and ergodic theories~\cite{Hinsley&Scully&Shilnikov2024,Araujo+2009}, they have not yet been utilized as a surrogate model and a straightforward parameter estimator for predicting extreme events.
The PRV map is considered universal across a wide range of extreme events arising from homoclinic bursting.
Thus, understanding the statistical properties of the PRV map may aid in developing a universal prediction theory for extreme events.

This study builds on a theoretical method for infinite-modal maps of this class, referred to as the randomization theory of infinite-modal maps. 
Previous studies~\cite{Nakagawa+2014,Nakagawa2015,Nakagawa-thesis} developed this method, analyzed several one-dimensional infinite-modal maps, and identified heavy-tailed invariant densities and distinctive burst statistics.
This analysis will provide a solid foundation for this study. 
This study also presents a parameter estimation method based on the theoretical results of the PRV map, which directly ties model parameters to observable statistics.
This can later help construct prediction methods that effectively utilize limited observations.

The remainder of this paper is organized as follows: 
Section \ref{sec:prv-map} describes the derivation of the PRV map.
Section \ref{sec:results_nume} presents the numerical results, including bifurcation diagrams, Lyapunov exponents, height probability distributions, and interevent interval probability distributions, for the PRV map.
Section \ref{sec:results_theo} presents the theoretical results, including the introduction of the randomized PRV map for analysis, as well as the analytical expressions for the mean, variance, and the stationary probability distribution of the map.
Section \ref{sec:para-estim} proposes a parameter estimation method for the PRV map based on the above theoretical results, including the parameter estimation procedure and a potential application for non-stationary data.
Section \ref{sec:concl} concludes the study and provides a scope for future studies.

\section{The PRV map}\label{sec:prv-map}
This preliminary section presents a derivation of the PRV map, following previous studies~\cite{Pacifico+1998,Shilnikov&Shilnikov2007}.

Before the derivation, a brief overview is provided, leading to the PRV map.
Shilnikov proved that a countably infinite number of periodic orbits exist near a homoclinic orbit to a saddle-focus point under certain conditions~\cite{Shilnikov1965}.
This indicates the presence of potential chaos near a homoclinic orbit.
Simultaneously, Shilnikov showed that a two-dimensional infinite-modal map arises on a Poincar\'{e} section near the saddle-focus point.
Furthermore, Arneodo et al.\ derived a one-dimensional version of this infinite-modal map as a dynamical system on the half-real line with leak by taking a strong dissipative limit (i.e., the limit of zero area reduction rate)~\cite{Arneodo+1985}.
Finally, Pacifico et al.\ derived an infinite-modal map on an interval as a closed dynamical system~\cite{Pacifico+1998}, referred to as the {\it Arneodo-Pacifico (AP) map}. 
The AP map exhibits chaotic behaviors with positive Lyapunov exponents in a parameter region of positive measure~\cite{Pacifico+1998, Araujo+2009}.
A previous study has investigated the statistical properties of the AP map from the perspective of on-off intermittency~\cite{Nakagawa2015}.
In this study, a two-dimensional version of this AP map, referred to as the PRV map, will be analyzed.

Let us derive the PRV map.
Consider a saddle-focus point with a homoclinic orbit in a three-dimensional autonomous vector field, as shown in Fig.~\ref{fig:homoclinic-loop}.
We assume that the saddle-focus point locates the origin O.
Therefore, the dynamics near the saddle-focus point can be described as the following linear differential equation:
\begin{align}
    \begin{pmatrix}
    \dot{x} \\ \dot{y} \\ \dot{z}
    \end{pmatrix}
    =
    \begin{pmatrix}
        -\alpha & \omega & 0 \\
        -\omega & -\alpha & 0 \\
        0 & 0 & \beta
    \end{pmatrix}
    \begin{pmatrix}
        x \\ y \\ z
    \end{pmatrix},
    \label{eq:ldeq}
\end{align}
where $\alpha, \beta, \omega$ are positive, which makes the $z$ direction of the origin O along the unstable manifold and the $xy$ plane along the stable manifold.
We assume the Shilnikov condition $\alpha < \beta$, which implies potential chaos near the homoclinic orbit~\cite{Shilnikov1965,Shilnikov&Shilnikov2007}.

Consider two sections $P_0$, $P_1$ near the origin, which are located on the plane $y=0$ and $z=h\ (>0)$, respectively.
See Fig.~\ref{fig:homoclinic-loop} for the $P_0$, $P_1$.
Let us construct the PRV map $T_\text{PRV}$ as a Poincar\'{e} map $P_0 \rightarrow P_0$, following (1)--(4):

\noindent
\textbf{(1) Construction of the map $T_1: P_0\rightarrow P_1$.} 
Denote a departure point by $(x_0, 0, z_0)\in P_0$ and the corresponding destination point by $(x_1, y_1, h)\in P_1$.
Since the $z$-component $\dot{z}=\beta z$ in Eq.~(\ref{eq:ldeq}), the travel time $\tau$ from $P_0$ to $P_1$ is described as
\begin{equation}
    \tau = \frac{1}{\beta} \log\frac{h}{z_0}.
\end{equation}
Therefore, substituting $\tau$ for $t$ in the solution $(x(t),y(t),z(t))$ of Eq.~(\ref{eq:ldeq}), we obtain the map 
\begin{align}
    T_1:
    \begin{dcases}
        x_1 = x_0 \left(\frac{z_0}{h}\right)^a \cos\left[b\log\left(\frac{z_0}{h}\right)\right], \\
        y_1 = x_0 \left(\frac{z_0}{h}\right)^a \sin\left[b\log\left(\frac{z_0}{h}\right)\right],
    \end{dcases}
\end{align}
where $a=\alpha/\beta$ and $b=\omega/\beta$.
From the Shilnikov condition, we assume $a<1$.

\noindent
\textbf{(2) Construction of the map $T_2: P_1\rightarrow P_0$.} 
No singular dynamics occur near the homoclinic orbit far from the origin.
Therefore, the map $T_2$ is a slightly deformed transformation, i.e., a combination of translation and rotation.
Denoting a departure point by $(x_1, y_1, h)\in P_1$ and the corresponding destination point by $(x_0, 0, z_0)\in P_0$, the map can be expressed as
\begin{align}
    T_2:
    \begin{pmatrix}
        x_0 \\
        z_0
    \end{pmatrix}
    =
    \begin{pmatrix}
        \tilde{x} & 0 \\
        0 & \tilde{z}
    \end{pmatrix}
    +
    \begin{pmatrix}
        \cos\phi & -\sin\phi \\
        \sin\phi & \cos\phi
    \end{pmatrix}
    \begin{pmatrix}
        x_1 \\
        y_1
    \end{pmatrix}.
\end{align}

\noindent
\textbf{(3) Construction of the map $T_3: P_0\rightarrow P_0$.} 
From the above, the map $T_3$ can be derived as $T_2\circ T_1$.
Denoting a departure point by $(x,0,z)\in P_0$ and the corresponding destination point by $(x',0,z')\in P_0$, the map can be expressed as
\begin{align}
    T_3=T_2\circ T_1:
    \begin{dcases}
        x' = x \left(\frac{z}{h}\right)^a \cos\left[b\log\left(\frac{z}{h}\right)+\phi\right]+\tilde{x}, \\
        z' = x \left(\frac{z}{h}\right)^a \sin\left[b\log\left(\frac{z}{h}\right)+\phi\right]+\tilde{z},
    \end{dcases}
\end{align}
The map $T_3:P_0\rightarrow P_0$ is defined only for $z>0$; thus, we do not consider the orbit on $z<0$.

\noindent
\textbf{(4) Construction of the PRV map $T_\text{PRV}$.}
Pacifico et al.~\cite{Pacifico+1998} extended the above `unclosed' dynamical system $T_3$ to the following closed dynamical system, the PRV map $T_\text{PRV}$:
\begin{align}
    T_\text{PRV} &: \mathbf{R}^2\rightarrow\mathbf{R}^2 \nonumber \\
    &
    \begin{dcases}
        x' = x \left(\frac{\lvert z\rvert}{h}\right)^a \cos\left[b\log\left(\frac{\lvert z\rvert}{h}\right)+\phi\right]+\tilde{x}, \\
        z' = \mathrm{sgn}\left(z\right) x \left(\frac{\lvert z\rvert}{h}\right)^a \sin\left[b\log\left(\frac{\lvert z\rvert}{h}\right)+\phi\right]+\tilde{z},
    \end{dcases}
    \label{eq:PRV}
\end{align}
where $a\in (0,1)$ and $b\in (0,\infty)$ are main parameters, and $h, \phi, \tilde{x}, \tilde{z}$ are sub parameters fixed in this study: $h=1$, $\phi=0$, $\tilde{x}=0.5$, and $\tilde{z}=0$.
The main parameter $b$ is also often fixed at a large value, e.g., $b = 100$, because these values do not significantly alter the statistical properties of the PRV map.

In the following sections, the statistical properties of the time series of the PRV map $x_{n+1} = T_\text{PRV}\left(x_n\right)$ will be presented, and often analyzed through the radial variable:
\begin{equation}
    r_n := \sqrt{(x_n-\tilde{x})^2+(z_n-\tilde{z})^2}.
\end{equation}

\section{Numerical results}\label{sec:results_nume}
This section presents numerical results for the statistical properties of the PRV map.
These include bifurcation diagrams, Lyapunov exponents, and height probability distributions of the spatial variables $x_n, z_n, r_n$; as well as probability distributions of interevent intervals with respect to laminar duration $n$ satisfying $r_n \leq r_\text{th}$.

\begin{figure}[tb]
    \centering
    \includegraphics[width=\linewidth]{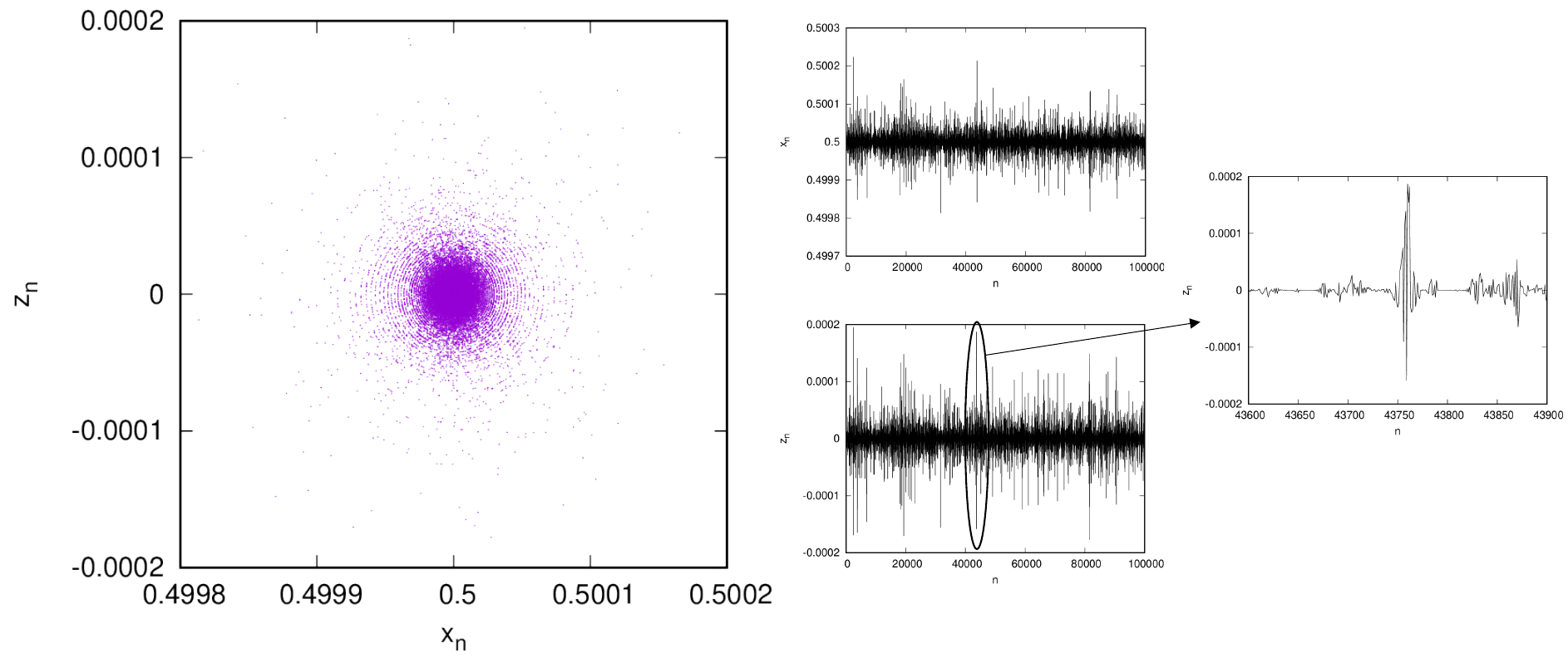}
    \caption{Typical orbit on $xz$ plane (left), the corresponding time series of $x_n$ (center top) and $z_n$ (center bottom), and the close-up of a typical burst (right). 
    The time series is generated from the PRV map with $a=0.9$, $b=100$, $h=1$, $\phi=0$, $\tilde{x}=0.5$, $\tilde{z}=0$.}
    \label{fig:PRV-map_orbit}
\end{figure}

\subsection{Bifurcation diagrams}\label{subsec:bif}
\begin{figure}[tb]
    \centering
    \includegraphics[width=\linewidth]{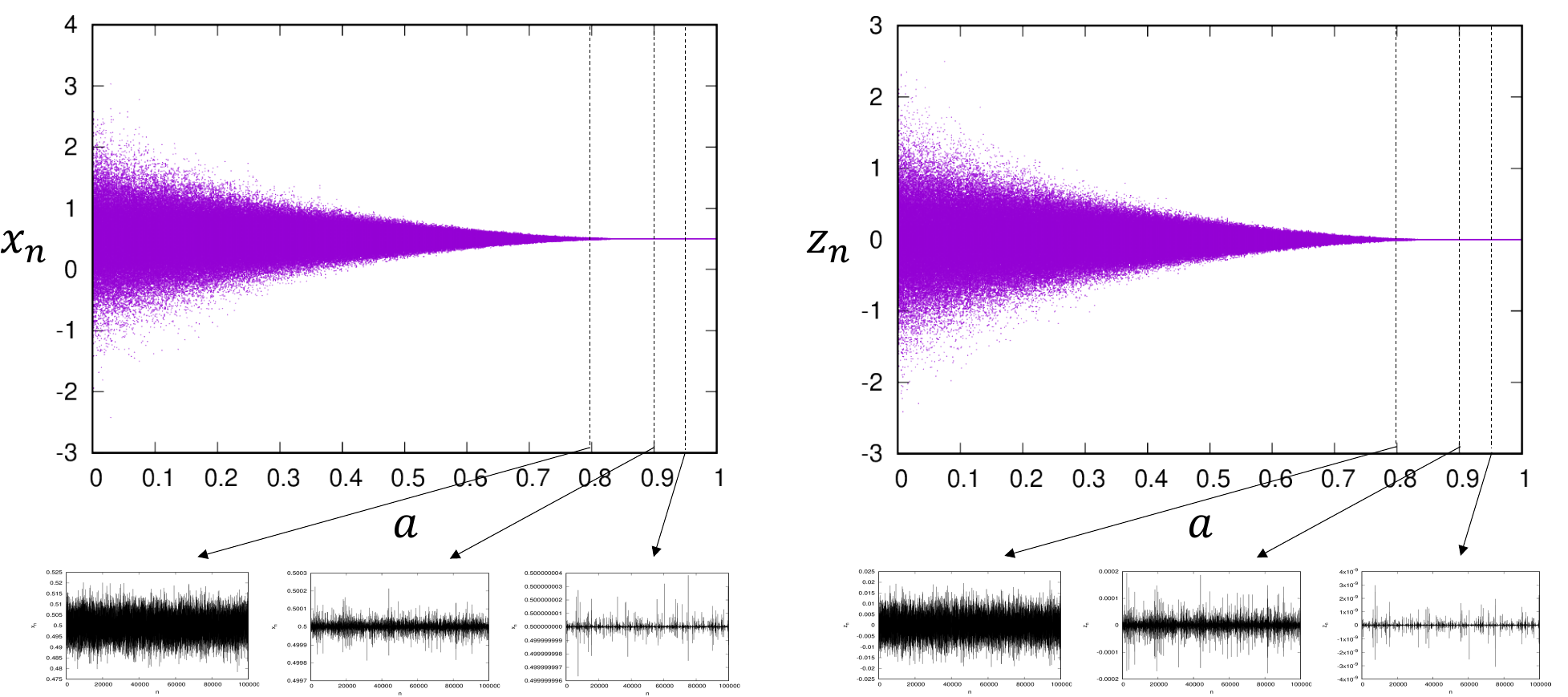}
    \caption{Bifurcation diagrams of the PRV map against the parameter $a\in(0,1)$ for the variables $x_n$ (left) and $z_n$ (right).
    The diagrams below illustrate various corresponding time series $x_n$ and $z_n$.
    Intermittent bursting intensifies as the parameter $a$ approaches $1$.
    Other parameters are fixed at $b=100$, $h=1$, $\phi=0$, $\tilde{x}=0.5$, $\tilde{z}=0$.}
    \label{fig:bif-diag}
\end{figure}

\begin{figure}[tb]
    \centering
    \includegraphics[width=.49\linewidth]{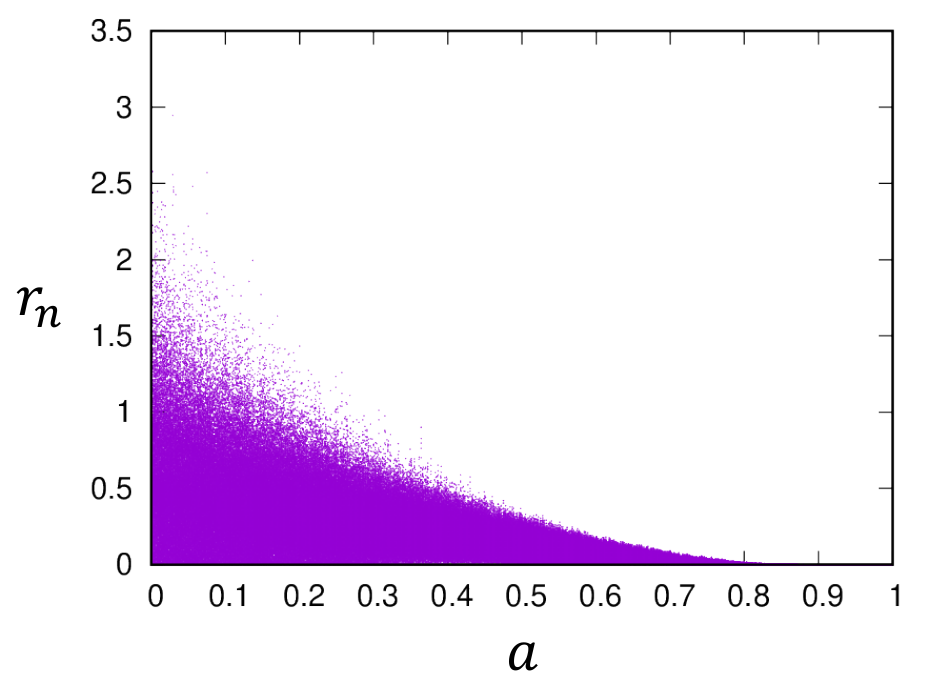}    
    \caption{Bifurcation diagrams of the PRV map against the parameter $a\in(0,1)$ for the radial variable $r_n$.
    The property of $x$ and $z$ is inherited by the radial variable $r$.
    Other parameters are fixed at $b=100$, $h=1$, $\phi=0$, $\tilde{x}=0.5$, $\tilde{z}=0$.}
    \label{fig:bif-diag_r}
\end{figure}

Figure \ref{fig:PRV-map_orbit} shows a typical orbit of the PRV map on the $xz$ plane (left), the corresponding time series of $x_n$ (center top) and $z_n$ (center bottom), and the close-up of a typical burst (right).
The time series exhibits intermittent bursting, corresponding to extreme events.
Figure \ref{fig:bif-diag} shows bifurcation diagrams of the PRV map against the main parameter $a\in(0,1)$ for the variables $x_n$ and $z_n$, respectively.
The bifurcation diagrams show that the heights $|x_n|$ and $|z_n|$ decrease, while the intermittency intensifies as the parameter $a$ approaches $1$.
Figure \ref{fig:bif-diag_r} shows the bifurcation diagram of the radial variable $r_n$ against the parameter $a\in(0,1)$.
Figures \ref{fig:bif-diag} and \ref{fig:bif-diag_r} indicate that the variables $x_n$ and $z_n$ have almost the same bifurcation diagram; moreover, their property is inherited by the radial variable $r_n$.

\subsection{Lyapunov exponents}\label{subsec:lyap}
\begin{figure}[tb]
    \centering
    \begin{minipage}[b]{.49\linewidth}
        \includegraphics[width=\linewidth]{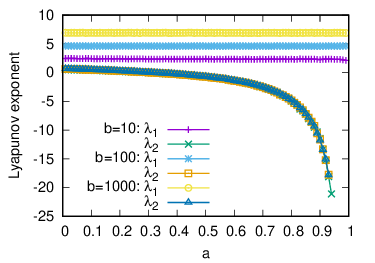}
    \end{minipage}
    \hfill
    \begin{minipage}[b]{.49\linewidth}
        \includegraphics[width=\linewidth]{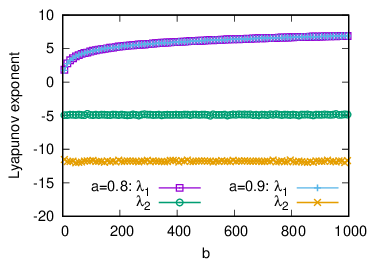}
    \end{minipage}
    \caption{Dependencies of Lyapunov exponents of the PRV map on the parameters $a$ (left) and $b$ (right). 
    Here, $\lambda_1$ and $\lambda_2$ denote the first (largest) and second Lyapunov exponents, respectively.
    Other parameters are fixed at $h=1$, $\phi=0$, $\tilde{x}=0.5$, $\tilde{z}=0$.}
    \label{fig:lyapunov}
\end{figure}

Figure \ref{fig:lyapunov} shows the dependencies of Lyapunov exponents for the PRV map on its parameters $a$ and $b$.
The first (largest) Lyapunov exponent $\lambda_1$ does not depend on the parameter $a$, but on the parameter $b$, i.e., as $b$ increases, $\lambda_1$ increases.
On the other hand, the second Lyapunov exponent $\lambda_2$ does not depend on the parameter $b$, but on the parameter $a$, i.e. as $a$ increases, $\lambda_2$ decreases.
There is at least one positive Lyapunov exponent for a wide range of parameters, which indicates that chaos indeed occurs.

\subsection{Height probability distributions}\label{subsec:h-dist}
\begin{figure}[tb]
    \centering
    \includegraphics[width=\linewidth]{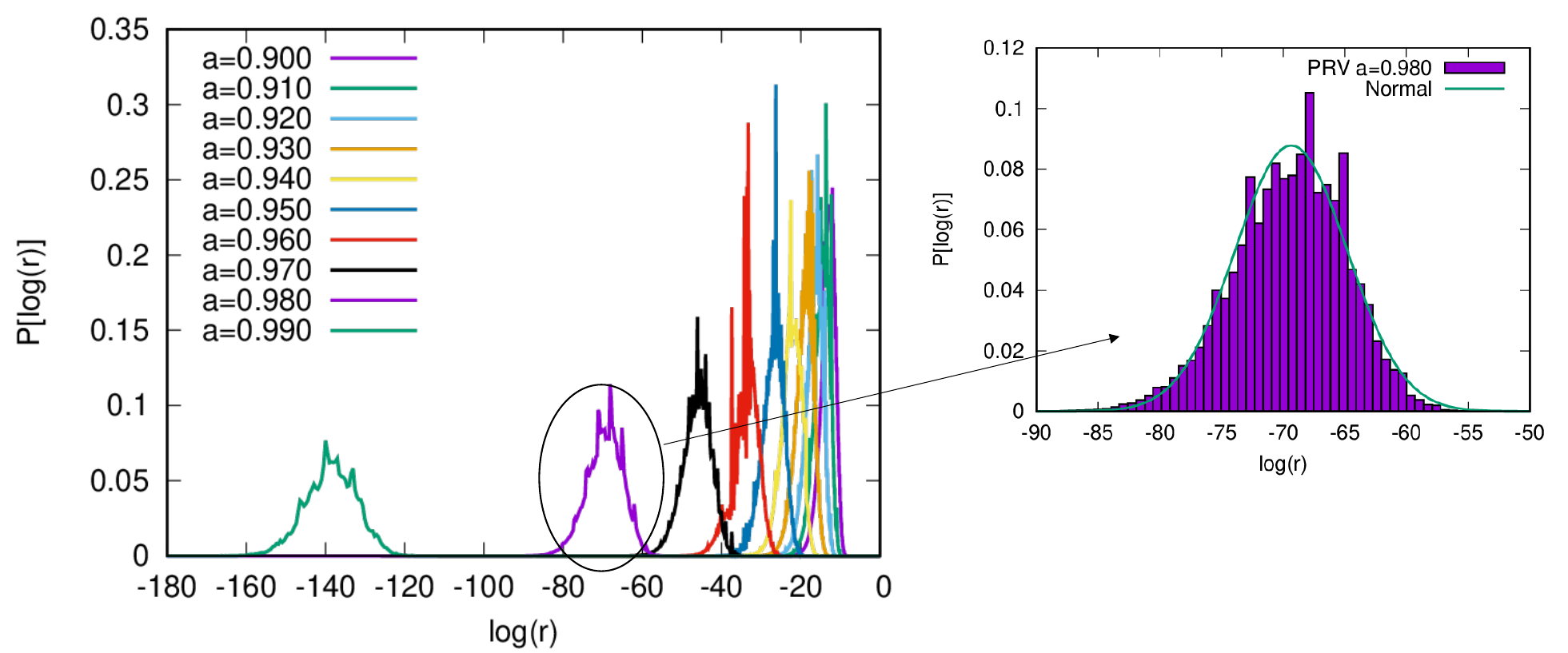}
    \caption{Histograms of logarithmic heights $\log{r_n}$ of the PRV map with $a$ near $1$.
    The right panel compares the histogram of $a=0.98$ with the normal distribution.
    Other parameters are fixed at $b=100$, $h=1$, $\phi=0$, $\tilde{x}=0.5$, $\tilde{z}=0$.}
    \label{fig:event-height-density}
\end{figure}

\begin{figure}[tb]
    \centering
    \begin{minipage}[b]{.49\linewidth}
        \includegraphics[width=\linewidth]{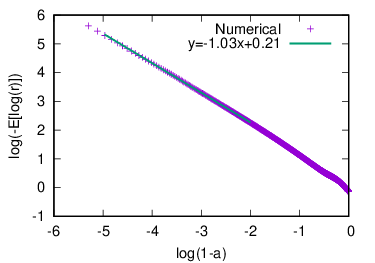}
    \end{minipage}
    \hfill
    \begin{minipage}[b]{.49\linewidth}
        \includegraphics[width=\linewidth]{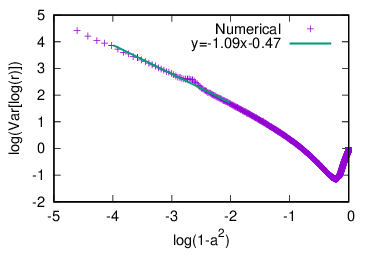}
    \end{minipage}
    \caption{Dependence of the mean (left) and the variance (right) for $\log{r_n}$ of the PRV map on the parameter $a$.
    The fitted lines are calculated on the range, as shown in the figure.
    Other parameters are fixed at $b=100$, $h=1$, $\phi=0$, $\tilde{x}=0.5$, $\tilde{z}=0$.}
    \label{fig:log-ave-var}
\end{figure}

We focus on the properties of the radial variable $r_n$.
Figure \ref{fig:event-height-density} shows histograms of logarithmic heights $\log{r_n}$ of the PRV map when $a$ is near $1$.
The right panel of Fig.~\ref{fig:event-height-density} compares the histogram with the fitted normal distribution.
Therefore, the probability distributions of the radial variable $r_n$ are approximated by the log-normal distributions with different means and variances, which mainly depend on the parameter $a$.
Figure \ref{fig:log-ave-var} shows the dependence of the mean and variance for $\log{r_n}$ on the parameter $a$.
The left panel of Fig.~\ref{fig:log-ave-var} shows that the mean for $\log{r_n}$ depend on the parameter $a$, with $\mathrm{E}[\log{r_n}]\propto (1-a)^{-1}$ as $a$ approaches $1$.
Similarly, the right panel of Fig.~\ref{fig:log-ave-var} shows that the variance for $\log{r_n}$ depend on the parameter $a$, with $\mathrm{Var}[\log{r_n}]\propto (1-a^2)^{-1}$ as $a$ approaches $1$.
Thus, the mean and variance have a clear dependence on the main parameter $a$ of intermittency.

\subsection{Interevent interval probability distributions}\label{subsec:iei-dist}
\begin{figure}[tb]
    \centering
    \begin{minipage}[b]{.49\linewidth}
        \includegraphics[width=\linewidth]{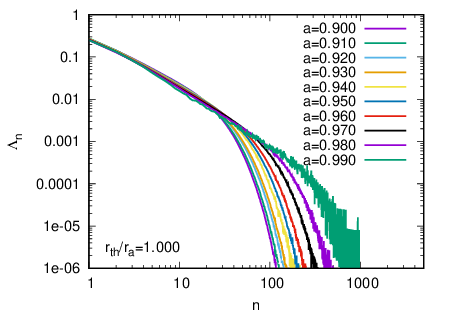}
    \end{minipage}
    \hfill
    \begin{minipage}[b]{.49\linewidth}
        \includegraphics[width=\linewidth]{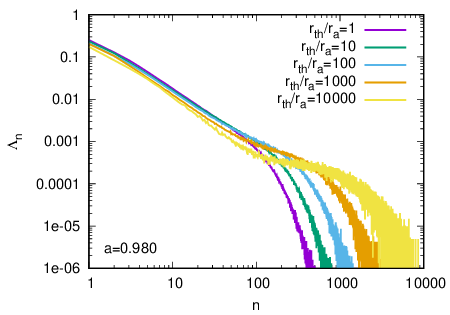}
    \end{minipage}
    \caption{Histograms of interevent intervals for $r_n$ of the PRV map with cases of the fixed event threshold $r_\text{th}=r_a$ (left) and the fixed main parameter $a=0.98$ (right).
    Other parameters are fixed at $b=100$, $h=1$, $\phi=0$, $\tilde{x}=0.5$, $\tilde{z}=0$.}
    \label{fig:laminar}
\end{figure}

The probability distribution of interevent intervals is defined as follows:
\begin{align}
    \Lambda_n = \mathrm{Prob.}\left\{r_1 \leq r_\text{th},\cdots,r_n \leq r_\text{th},r_{n+1} > r_\text{th} \mid r_0 > r_\text{th}\right\},
\end{align}
where $r_\text{th}$ is a threshold for determining extreme events.
Figure \ref{fig:laminar} shows histograms of interevent intervals for $r_n$ of the PRV map with $a$ near $1$.
An event occurs when the radial variable $r_n$ crosses over a threshold $r_\mathrm{th}$, which is set to several times the reference value
\begin{align}
    r_a := \exp\left\{\mathrm{E}\left[\log r_n\right]\right\}.
\end{align}
The left panel of Fig.~\ref{fig:laminar} shows histograms for different parameters $a$ with the fixed threshold $r_\mathrm{th}=r_a$.
The right panel of Fig.~\ref{fig:laminar} shows histograms for different thresholds $r_\mathrm{th}$ with the fixed parameter $a=0.98$.
According to the left panel of Fig.~\ref{fig:laminar}, if the threshold $r_\mathrm{th}$ is near $r_a$, the probability distribution of interevent intervals exhibits a power law in the short-time region, followed by exponential decay in the long-time region.
However, according to the right panel of Fig.~\ref{fig:laminar}, if the threshold $r_\mathrm{th}$ is significantly above $r_a$, the probability distribution of interevent intervals develops a `shoulder' in the long-time region before exhibiting exponential decay.

\begin{figure}[tb]
    \centering
    \begin{minipage}[b]{.49\linewidth}
        \includegraphics[width=\linewidth]{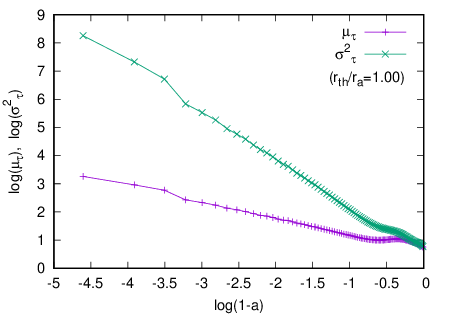}
    \end{minipage}
    \hfill
    \begin{minipage}[b]{.49\linewidth}
        \includegraphics[width=\linewidth]{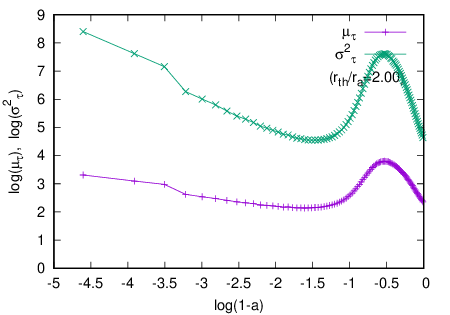}
    \end{minipage}
    \\
    \begin{minipage}[b]{.49\linewidth}
        \includegraphics[width=\linewidth]{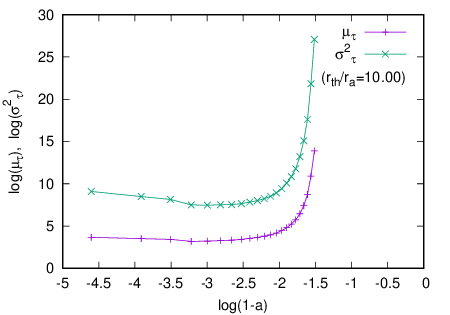}
    \end{minipage}
    \hfill
    \begin{minipage}[b]{.49\linewidth}
        \includegraphics[width=\linewidth]{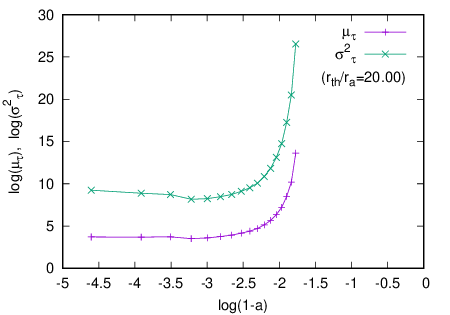}
    \end{minipage}
    \caption{Dependencies of the mean $\mu_\tau$ and variance $\sigma^2_\tau$ of interevent intervals for $r_n$ of the PRV map against the parameter $a\in (0,1)$: the event thresholds $r_\text{th}/r_a = 1$, $2$, $10$, $20$ (left to right, and top to bottom).
    Other parameters are fixed at $b=100$, $h=1$, $\phi=0$, $\tilde{x}=0.5$, $\tilde{z}=0$.}
    \label{fig:laminar_ave-var}
\end{figure}

Figure \ref{fig:laminar_ave-var} shows dependencies of the mean $\mu_\tau$ and variance $\sigma^2_\tau$ of interevent intervals for $r_n$ of the PRV map against the parameter $a\in (0,1)$.
The thresholds in Fig.~\ref{fig:laminar_ave-var} are set to $r_\text{th}/r_a = 1$, $2$, $10$, $20$ (left to right, and top to bottom).
The mean and variance of interevent intervals exhibit the same dependency across the entire parameter $a\in (0,1)$.
The dependency is linear for $a$ near $1$.
Moreover, if the threshold is sufficiently above $r_a$, the mean and variance of interevent intervals become almost constant for $a$ near $1$.
Thus, the probability distribution of interevent intervals exhibits somewhat complex dependence on the parameter $a$ and the threshold $r_\text{th}$.

\section{Theoretical results}\label{sec:results_theo}
This section presents theoretical results for the statistical properties of the PRV map, including the analytical formulae for the mean, variance, and the stationary probability distribution (equivalent to the height probability distribution) of $\log r_n$.
The theoretical results will be derived by analyzing a randomized version of the PRV map, which is based on the randomization theory of infinite-modal maps~\cite{Nakagawa-thesis,Nakagawa2015,Nakagawa+2014}.

\subsection{Randomized PRV map}\label{subsec:r-prv-map}
\begin{figure}[tb]
    \centering
    \includegraphics[width=.49\linewidth]{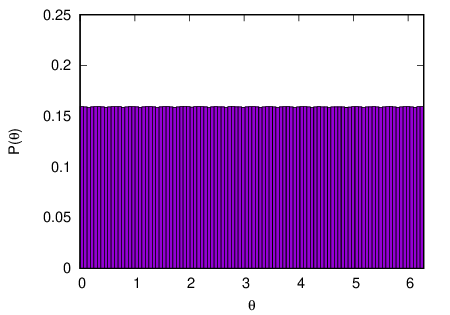}
    \caption{Histogram of the angular coordinate $\tilde{\theta}_n:= b \log\left(\frac{|z_n|}{h}\right)+\phi \mod 2\pi$ generated from a time series $(x_n,z_n)$ of the PRV map with $a=0.9$, $b=100$, $h=1$, $\phi=0$, $\tilde{x}=0.5$, $\tilde{z}=0$. 
    This is approximated by a uniform distribution on $[0,2\pi)$, which supports the uniform distribution hypothesis.}
    \label{fig:phase-density}
\end{figure}

According to the randomization theory~\cite{Nakagawa-thesis,Nakagawa2015,Nakagawa+2014}, a dynamical system with infinitely many critical points can be transformed into a random dynamical system by replacing the angular coordinate of a periodic function with a uniform random variable.
This assumption is known as the uniform distribution hypothesis~\cite{Nakagawa+2014,Nakagawa-thesis}.
Thus, we obtain from the PRV map~(\ref{eq:PRV}), the following randomized PRV (R-PRV) map $T_\text{R-PRV}$:
\begin{align}
    T_\text{R-PRV} &: \mathbf{R}^2\rightarrow\mathbf{R}^2 \nonumber \\
    &
    \begin{dcases}
        x_{n+1} = x_n \left(\frac{|z_n|}{h}\right)^a \cos\left(\theta_n \right)+\tilde{x},\\
        z_{n+1} = \mathrm{sgn}\left(z_n\right) x \left(\frac{|z_n|}{h}\right)^a \sin\left(\theta_n \right)+\tilde{z},
    \end{dcases}
    \label{eq:rand-PRV}
\end{align}
where $\{\theta_n\}$ are independent and identically distributed random variables that follow a uniform distribution on $[0,2\pi)$.
The histogram of the angular coordinate $\tilde{\theta}_n:= b \log\left(\frac{|z_n|}{h}\right)+\phi \mod 2\pi$ calculated from an actual time series $(x_n,z_n)$ of the PRV map is shown in Fig.~\ref{fig:phase-density}.
The histogram can indeed be approximated by the uniform distribution on $[0,2\pi)$, following the uniform distribution hypothesis.

The R-PRV map (\ref{eq:rand-PRV}) can be translated into a polar coordinate system $(r_n,\theta_n)$ by a transformation $r_n := \sqrt{(x_n-\tilde{x})^2+(z_n-\tilde{z})^2}$, $\theta_n := \mathrm{atan2}\left(z_n-\tilde{z},x_n-\tilde{x}\right)$:
\begin{equation}
    r_{n+1} = \left|\tilde{x}+r_n\cos{\theta_n}\right| \left(\frac{\left|\tilde{z}+r_n\sin{\theta_n}\right|}{h}\right)^a. 
\end{equation}
Assuming $\tilde{z}\ll r_n$ and $r_n\ll \tilde{x}$, and denoting $c:=h^{-a}|\tilde{x}|$ and $\xi_n:=|\sin\varphi_n|$, we obtain the following ``radial map:''
\begin{equation}
    r_{n+1} = c {\xi_n}^a {r_n}^a. \label{eq:radial-map}
\end{equation}
$\{\xi_n\}$ follow the following probability density function:
\begin{equation}
    \rho(\xi)=\frac{2}{\pi\sqrt{1-\xi^2}}\quad (\xi\in[0,1]).
\end{equation}
Therefore, we focus on analyzing either the radial map (\ref{eq:radial-map}) or the following ``logarithmic radial map:''
\begin{equation}
    w_{n+1} = a w_n + a \eta_n - \log c, 
    \label{eq:log-radial-map}
\end{equation}
where $w_n := -\log r_n$ and $\eta_n := -\log\xi_n$.
The corresponding logarithmic radial map for the one-dimensional AP map is known to take a similar form: $w_{n+1} = aw_n + \eta_n - \log c$~\cite{Nakagawa2015}.

\subsection{Mean and variance of $\{w_n\}$}\label{subsec:ave-var-w}
First, we derive the mean of the logarithmic radial map $\{w_n\}$, assuming the random variables $w_n$ and $\eta_n$ are stationary.
Averaging both sides of Eq.~(\ref{eq:log-radial-map}) and denoting $\mu := \mathrm{E}[w_n]$ and $\mu_0 := \mathrm{E}[\eta_n]$, we obtain $\mu = a \mu + a \mu_0 - \log c$.
Therefore, using $\mu_0 = \log 2$, the mean $\mu$ of $\{w_n\}$ is derived:
\begin{align}
    \mu &= \frac{a\log 2 - \log c}{1-a} \nonumber\\
    &=\frac{a\log(2h)-\log|\tilde{x}|}{1-a}. 
    \label{eq:ave_w}
\end{align}
The corresponding mean for the one-dimensional AP map is known to have a slightly different form: $\mu = \frac{\log 2 - \log c}{1-a}$~\cite{Nakagawa2015}.

Next, we derive the variance of the logarithmic radial map $\{w_n\}$.
Subtracting $\mu := \mathrm{E}[w]$ from both sides of Eq.~(\ref{eq:log-radial-map}), squaring the results, and averaging, while using the relation $(1-a)\mu = a\mu_0 - \log c$, we obtain:
\begin{align}
    \mathrm{E}[(w_{n+1}-\mu)^2] 
    &= \mathrm{E}[(a w_n + a \eta_n-\log{c}-\mu)^2] \nonumber \\
    &= \mathrm{E}[\{a(w_n-\mu)+a(\eta_n-\mu_0)\}^2] \nonumber \\
    &= a^2\mathrm{E}[(w_n-\mu)^2] + a^2\mathrm{E}[(\eta_n-\mu_0)^2],
\end{align}
where $\mathrm{E}[(w_n-\mu)(\eta_n-\mu_0)]=0$ is assumed, i.e., no correlation between $(w_n-\mu)$ and $(\eta_n-\mu_0)$.
Denoting $\sigma^2 := \mathrm{E}[(w_n-\mu)^2]$ and $\sigma_0^2 := \mathrm{E}[(\eta_n-\mu_0)^2]$, and using $\sigma_0^2 = \pi/\sqrt{12}$, the variance $\sigma^2$ of $\{w_n\}$ is derived:
\begin{equation}
    \sigma^2 = \frac{a^2}{1-a^2}\frac{\pi}{\sqrt{12}}. \label{eq:var_w}
\end{equation}
The corresponding variance for the one-dimensional AP map is known to have a slightly different form: $\sigma^2 = \frac{1}{1-a^2}\frac{\pi}{\sqrt{12}}$~\cite{Nakagawa2015}.

\begin{figure}[tb]
    \centering
    \begin{minipage}[b]{.49\linewidth}
        \includegraphics[width=\linewidth]{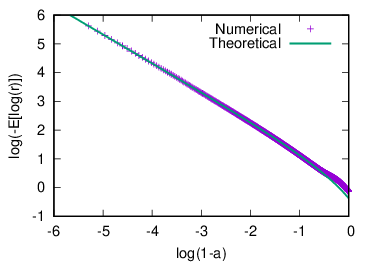}
    \end{minipage}
    \hfill
    \begin{minipage}[b]{.49\linewidth}
        \includegraphics[width=\linewidth]{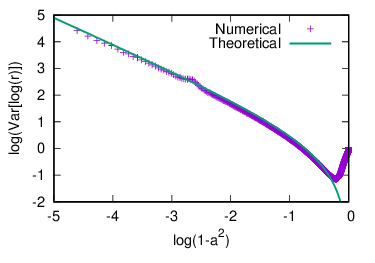}
    \end{minipage}    
    \caption{
    Dependence of the mean (left) and the variance (right) for $\{w_n\}$ on the parameter $a$.
    The numerical points are calculated from the PRV map (\ref{eq:PRV}) with $b=100$, $h=1$, $\phi=0$, $\tilde{x}=0.5$, $\tilde{z}=0$, and the theoretical line is drawn by Eq.~(\ref{eq:ave_w}).}
    \label{fig:ave-var_w-theory}
\end{figure}

Figure \ref{fig:ave-var_w-theory} compares theoretical means and variances given by Eqs.~(\ref{eq:ave_w}), (\ref{eq:var_w}) and numerical values calculated from the PRV map (\ref{eq:PRV}).
The theoretical means are consistent with the numerical values, and the theoretical variances are approximately consistent with the numerical values.
The variance (\ref{eq:var_w}) depends only on the parameter $a$, which will be utilized in the following parameter estimation method.

\subsection{Stationary probability distribution of $\{w_n\}$}\label{subsec:w-dist}
\begin{figure}[tb]
    \centering
    \includegraphics[width=.75\linewidth]{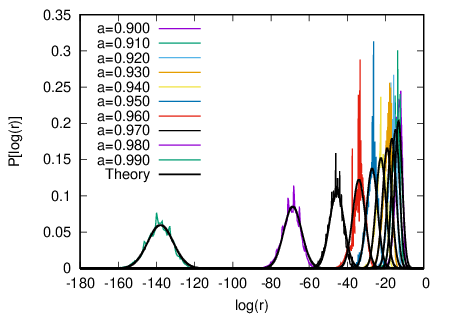}
    \caption{Stationary probability distributions of logarithmic event heights $\{\log{|r_n|}\}$ of the PRV map with $a$ near $1$.
    Theoretical lines (black) are drawn by Eq.~(\ref{eq:den_w}), and numerical lines (colored) are calculated from the PRV map (\ref{eq:PRV}) with $b=100$, $h=1$, $\phi=0$, $\tilde{x}=0.5$, $\tilde{z}=0$.}
    \label{fig:denl_w-theory}
\end{figure}

We derive the stationary probability distribution of $\{w_n\}$, which corresponds to the height probability distribution for $\log r_n$ of the PRV map.
First, considering $\frac{w_n-\mu}{\sigma}$ in Eq.~(\ref{eq:log-radial-map}), we obtain the following standardized logarithmic radial map:
\begin{equation}
    \frac{w_{n+1}-\mu}{\sigma} = a \frac{w_n-\mu}{\sigma} + \sqrt{1-a^2} \,\frac{\eta_n-\eta_0}{\sigma_0}.
\end{equation}
Next, considering the characteristic functions $\psi_n(t) := \mathrm{E}\left[\exp\left(it\frac{w_n-\mu}{\sigma}\right)\right]$ and $\varphi_n(t) := \mathrm{E}\left[\exp\left(it\frac{\eta_n-\mu_0}{\sigma_0}\right)\right]$, these satisfy the following functional relation:
\begin{equation}
    \psi_{n+1}(t) = \psi_n(at) \varphi_n\left(\sqrt{1-a^2}\,t\right). \label{eq:ch-funcs_relation}
\end{equation}
The functional relation (\ref{eq:ch-funcs_relation}) was investigated in a previous study.~\cite{Nakagawa2015}
According to the study, $\psi_n(t)$ goes to $\exp\left(-\frac{t^2}{2}\right)$ when $n\rightarrow \infty$ and $a\rightarrow 1$.
Therefore, $\left\{\frac{w_n-\mu}{\sigma}\right\}$ follow asymptotically the standard normal distribution when $a$ is near $1$.
Thus, the stationary probability distribution of $\{\log r_n\}\ (r_n = e^{-w_n})$ for $a\approx 1$ is a normal distribution with the mean $\mu$ and variance $\sigma^2$ given by Eqs.~(\ref{eq:ave_w}) and (\ref{eq:var_w}):
\begin{align}
    P[\log r] = \frac{1}{\sqrt{2\pi \sigma^2}} \exp\left[-\frac{(\log r + \mu)^2}{2\sigma^2}\right].
    \label{eq:den_w}
\end{align}
Therefore, the stationary probability distribution of the radial map $\{r_n\}$ is a log-normal distribution.
Figure \ref{fig:denl_w-theory} compares the stationary probability distributions of $\{\log r_n\}$ numerically calculated from the PRV map (\ref{eq:PRV}) with the theoretical one given by Eq.~(\ref{eq:den_w}).
The comparison shows close agreement.
The stationary probability distribution of the one-dimensional AP map is also known to have the log-normal distribution~\cite{Nakagawa2015}.

The statistical properties of the PRV map are nearly identical to those of the one-dimensional AP map.
However, when the parameter $a$ deviates significantly from $1$, the mean and variance differ from those of the AP map.

\section{Parameter estimation of the PRV map}\label{sec:para-estim}
This section presents a parameter estimation method utilizing the analytical expressions obtained in the previous section.
The target to be estimated is the parameter $a$, which is a primary parameter for intermittency.

\subsection{The parameter estimation method}\label{subsec:para-estim-proce}
\begin{figure}[tb]
    \centering
    \begin{minipage}[b]{.49\linewidth}
        \includegraphics[width=\linewidth]{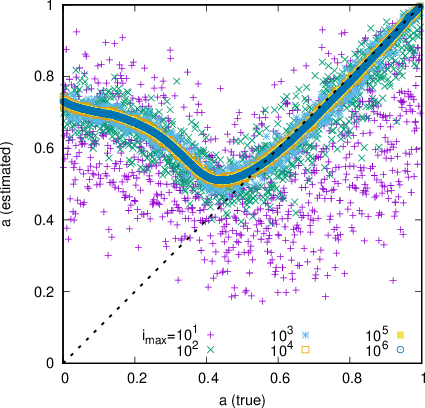}    
    \end{minipage}
    \hfill
    \begin{minipage}[b]{.49\linewidth}
        \includegraphics[width=\linewidth]{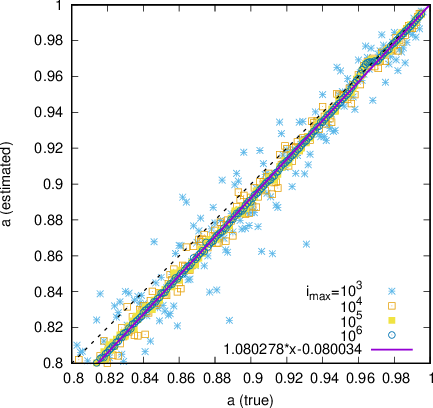}
    \end{minipage}
    \caption{Estimation of the parameter $a$ for the PRV map using the formula (\ref{eq:a-vs-sigma2}) with $b=100$, $h=1$, $\phi=0$, $\tilde{x}=0.5$, $\tilde{z}=0$.
    The figures represent the range of the whole $0<a<1$ (left) and the enlarged range $0.8\leq a<1$ (right).
    The data length used for the estimation is denoted by $i_{\rm max}$, ranging from $10$ to $10^6$.
    The variance is estimated by the unbiased variance from a time series $\{w_{n-i_\text{max}}, \cdots, w_n\}$.}
    \label{fig:pred-a}
\end{figure}

The variance formula (\ref{eq:var_w}) depends solely on the parameter $a$ and is one-to-one correspondence with it:
\begin{equation}
    a = \left(1+\frac{\pi}{\sqrt{12}\,\sigma^2}\right)^{-1/2}. \label{eq:a-vs-sigma2}
\end{equation}
Thereby, we can estimate the parameter $a$ from the variance $\sigma^2$ of the observed time series $\{w_n\}\ (w_n=-\log r_n)$.
Figure \ref{fig:pred-a} shows the correspondence between the actual value $a\text{(true)}$ and the estimated value $a\text{(estimated)}$ using Eq.~(\ref{eq:a-vs-sigma2}).
The estimated value $a\text{(estimated)}$ approaches the actual value $a\text{(true)}$ when the data length in the estimation of the variance is large enough and $a\text{(true)}$ is near $1$, as shown in Fig.~\ref{fig:pred-a}(left).
Parameter estimation is possible even for limited observations if such a trade-off relationship between accuracy and data length for estimation is acceptable.
However, a slight deviation is observable between the estimated and actual values, even when $a\text{(true)}$ is near $1$ enough, as shown in Fig.~\ref{fig:pred-a}(right).
The deviation shows a linear dependence as follows:
\begin{equation}
    a_\text{estimated} \approx 1.08\, a_\text{true} - 0.08.
\end{equation}
The origin of this linear dependence is unclear; however, we employ it as an empirical correction formula.

Therefore, we propose the following parameter estimation procedure for the PRV map with unknown parameter $a\in (0,1)$:
\begin{enumerate}
    \item Calculate the (unbiased) variance $s^2$ from a past time series $\{w_{n-i_\text{max}}, \cdots, w_n\}$:
        \begin{align}
            s^2 = \frac{1}{i_\text{max}-1} \sum_{i=0}^{i_\text{max}} \left(w_{n-i} - m \right),
        \end{align}
    where $m$ is a sample mean: $m = \frac{1}{i_\text{max}} \sum_{i=0}^{i_\text{max}} w_{n-i}$.
    \item Estimate the parameter $\tilde{a}_\text{estimated}$ from the calculated variance $s^2$:
        \begin{align}
            \tilde{a}_\text{estimated} = \left(1+\frac{\pi}{\sqrt{12}\,s^2}\right)^{-1/2}.
        \end{align}
    \item Correct the parameter $\tilde{a}_\text{estimated}$ following the empirical correction formula:
        \begin{align}
            \tilde{a}_\text{corrected} = \frac{\tilde{a}_\text{estimated}+0.08}{1.08}.
        \end{align}
\end{enumerate}
Other parameters ($b,h,\phi,\tilde{x},\tilde{z}$) are unused in the above estimation procedure.

Here, we assume that the parameters $\tilde{x},\tilde{z}$ are known.
They can be determined in advance from the center coordinates of the variables $x_n, z_n$ of the PRV map, as shown in Fig.~\ref{fig:PRV-map_orbit}(left).
Furthermore, the parameters $b$ and $\phi$ are considered less critical, since the statistical properties of the PRV map remain practically unchanged for large $b$ and any $\phi$.
Therefore, only the parameters $(a,h)$ must be treated as effectively unknown.
The value of the parameter $h$ can be estimated by the analytical expression of the mean, Eq.~(\ref{eq:ave_w}), from a past time series $\{w_{n-i_\text{max}},\cdots,w_n\}$:
\begin{align}
    h_\text{estimated} = \frac{1}{2}\exp{\left[\frac{m(1-\tilde{a}_\text{corrected}) + \log|\tilde{x}|}{\tilde{a}_\text{corrected}}\right]}.
\end{align}
This allows all the effective parameters $(a,h)$ to be determined. 
Thus, samples can be generated through numerical simulations to investigate various statistical laws governing extreme events, such as the probability distribution of interevent intervals.

\subsection{Applicability of the parameter estimation method to non-stationary data}\label{subsec:appl-para-estim}
\begin{figure}[tb]
    \centering
    \begin{minipage}[b]{.49\linewidth}
        \includegraphics[width=\linewidth]{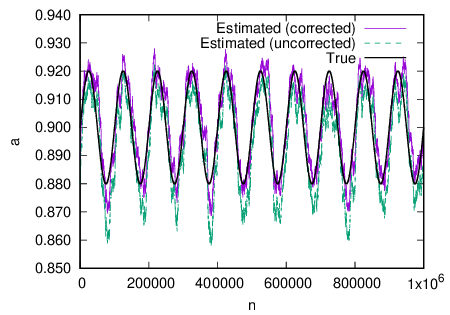}\\
        \includegraphics[width=\linewidth]{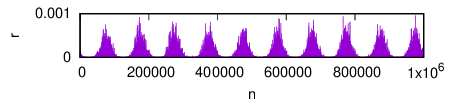}
    \end{minipage}
    \hfill
    \begin{minipage}[b]{.49\linewidth}
        \centering
        \includegraphics[width=\linewidth]{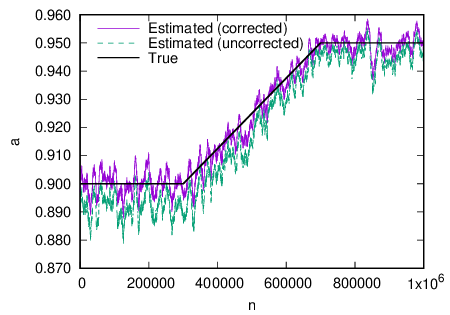}\\
        \includegraphics[width=\linewidth]{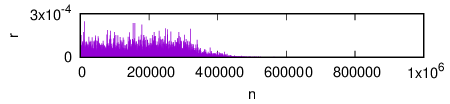}
    \end{minipage}
    \caption{Application examples of the parameter estimation method to non-stationary data for the PRV map with the time-dependent parameter $a$.
    Other parameters are fixed: $b=100$, $h=1$, $\phi=0$, $\tilde{x}=0.5$, $\tilde{z}=0$.
    The figures illustrate cases of a cyclic parameter (left column) and a linearly changing parameter (right column).
    The upper row compares the actual and estimated (uncorrected/corrected) parameters, {while} the lower row compares the time series used for parameter estimation of the PRV map.}
    \label{fig:pred-a_cycle-linear}
\end{figure}

This section presents the potential applicability of the proposed estimation method to non-stationary data.
Figure \ref{fig:pred-a_cycle-linear} shows two examples for such situations: a cyclic parameter case (left column) and a linearly changing parameter case (right column).
The cyclic parameter in the left column of Fig.~\ref{fig:pred-a_cycle-linear} is set to the following sinusoidal function:
\begin{align}
    a_n = A_0 + A_1 \sin\left(2\pi \frac{n}{T}\right),
\end{align}
where $A_0=0.90$, $A_1=0.02$, $T=10^5$.
The linearly changing parameter in the right column of Fig.~\ref{fig:pred-a_cycle-linear} is set to the following piecewise linear function:
\begin{align}
    a_n = 
    \begin{dcases}
        B_1 & (n<n_1), \\
        B_1 + (B_2-B_1)(n-n_1) & (n_1\leq n< n_2), \\
        B_2 & (n\geq n_2).
    \end{dcases}
\end{align}
where $B_1=0.90$, $B_2=0.95$, $n_1=3\times 10^5$, $n_2=7\times 10^5$.
In both cases, the data length used for the variance estimation is set to $i_\text{max}=10^4$.
The value of $i_\text{max}$ is significant for good results: smaller values of $i_\text{max}$ yield larger fluctuations, while larger values of $i_\text{max}$ yield over-averaged results.
Practically, setting an appropriate $i_\text{max}$ must be done through trial and error.

\section{Conclusions}\label{sec:concl}
This study analyzed homoclinic bursting through the PRV map as a mechanism-faithful but simple surrogate model.
Numerical results showed that height probability distributions approach log-normal distributions under strong intermittency, while probability distributions of interevent intervals additionally depend strongly on the event threshold.
Theoretical analysis using the R-PRV map yielded explicit formulae for mean, variance, and stationary distributions, and revealed close correspondence with the one-dimensional AP map.
Based on these results, a parameter estimation method was proposed and tested for data with time-dependent parameters, demonstrating potential applications to non-stationary data.

These findings provide a foundation for mechanism-based prediction of extreme events.
Future directions include refining the parameter estimation method.
They clarify its predictive utility, extending it to limited and short datasets, which is a critical challenge for practical forecasting.
This method may contribute to the development of strategies to mitigate extreme events driven by homoclinic bursting.

\acknowledgments 
The author would like to thank Dr.~Sho Shirasaka of the University of Osaka for his helpful comments.

\dataavailability
The data that support the findings of this study are available from the author upon reasonable request.

\funding
This study is supported by JSPS KAKENHI Grant No.~JP21K17825 and JST Moonshot R\&D Grant No.~JP-MJMS2021, Japan.

\conflictsofinterest
The authors declare no competing interests.

\authorcontribution
The author confirms sole responsibility for this work.

\bibliography{ref}
\bibliographystyle{unsrt}

\end{document}